\documentclass[conference,a4paper]{IEEEtran}

\usepackage{xcolor}
\usepackage{balance}
\usepackage{soul}
\usepackage{comment}
\usepackage{amssymb}
\ifCLASSINFOpdf
  \usepackage[pdftex]{graphicx}
\else
  \usepackage[dvips]{graphicx}
\fi
\ifCLASSOPTIONcompsoc
 \usepackage[caption=false,font=normalsize,labelfont=sf,textfont=sf]{subfig}
\else
 \usepackage[caption=false,font=footnotesize]{subfig}
\fi
\usepackage{amsmath}

\begin{document}
\title{Experimental Validation of SBFD ISAC in an FR3 Distributed SIMO Testbed}

\author{\IEEEauthorblockN{
Bixing Yan\IEEEauthorrefmark{1},   
Kwadwo Mensah Obeng Afrane\IEEEauthorrefmark{1},   
Achiel Colpaert\IEEEauthorrefmark{2},    
Andre Kokkeler\IEEEauthorrefmark{1},
Sofie Pollin\IEEEauthorrefmark{3},      
Yang Miao\IEEEauthorrefmark{1}
}                                     
\IEEEauthorblockA{\IEEEauthorrefmark{1}Radio Systems Group, University of Twente, Enschede, The Netherlands}
\IEEEauthorblockA{ \IEEEauthorrefmark{1}Emails:\{b.yan, k.m.obengafrane, a.b.j.kokkeler, y.miao\}@utwente.nl}
\IEEEauthorblockA{\IEEEauthorrefmark{2}Imec, Kapeldreef 75, 3001 Leuven, Belgium}
\IEEEauthorblockA{ \IEEEauthorrefmark{2}Emails:achiel.colpaert@imec.be}
\IEEEauthorblockA{\IEEEauthorrefmark{3}WaveCoRE, Department of Electrical Engineering (ESAT), KU Leuven, Belgium}
\IEEEauthorblockA{ \IEEEauthorrefmark{3}Emails:sofie.pollin@kuleuven.be}
}

\maketitle

\begin{abstract}
Integrated sensing and communication (ISAC) is a key enabler for future radio networks. This paper presents a sub-band full-duplex (SBFD) ISAC system that assigns non-overlapping OFDM subbands to sensing and communication, enabling simultaneous operation with minimal interference. A distributed testbed with three SIMO nodes is implemented using USRP~X410 devices operating at 6.8~GHz with 20~MHz bandwidth per channel. A total of 2048 OFDM subcarriers are partitioned into three subbands: two for sensing using Zadoff–Chu sequences and one for communication using QPSK. Each USRP transmits one subband while receiving signals across all three, forming a $1 \times 3$ SIMO node. Time synchronization is achieved through host-server coordination without external clock distribution.  
Indoor measurements, validated against MOCAP ground truth, confirm the feasibility of the SBFD ISAC system. The results demonstrate monostatic sensing with a velocity resolution of 0.145~m/s, and communication under NLoS conditions with a BER of $3.63 \times 10^{-3}$. Compared with a multiband benchmark requiring three times more spectrum, the SBFD configuration achieves comparable velocity estimation accuracy while conserving resources. The sensing and communication performance trade-off is determined by subcarrier allocation strategy rather than mutual interference.
\end{abstract}

\vskip0.5\baselineskip
\begin{IEEEkeywords}
Subband full-duplex, integrated sensing and communication, testbed and experiment, performance evaluation
\end{IEEEkeywords}

\section{Introduction}
Advances in radio network technologies have enabled the emergence of Integrated Sensing and Communication (ISAC), which unifies radar sensing and wireless communication within a single system. ISAC can be realized at varying integration levels by leveraging either shared or dedicated infrastructure, spectrum, waveforms, and time-frequency resources. A highly integrated approach uses shared orthogonal frequency-division multiplexing (OFDM) waveforms for both sensing and communication \cite{ofdm}. OFDM, already a cornerstone of modern wireless communications, has also proven effective for radar sensing \cite{ofdms}. To support concurrent operation, ISAC systems typically adopt time-division duplexing (TDD) or frequency-division duplexing (FDD) strategies \cite{tddfdd}. TDD-based ISAC alternates between sensing and communication within the same frequency band, conserving spectral resources but limiting continuous environmental sensing. By contrast, FDD-based ISAC assigns separate frequency bands to sensing and communication, ensuring continuous operation but at the cost of higher spectrum demand and additional radio hardware. 

Sub-band full-duplex (SBFD) technology enables simultaneous communication and sensing while optimizing spectrum usage and reducing hardware complexity in radio systems. Originally explored for full-duplex communication \cite{sbfd1,sbfds}, SBFD leverages frequency and time resource allocations by assigning distinct subcarriers within an OFDM waveform to uplink and downlink. Operating within a single frequency band, interference between functions is mitigated by interleaved guard subcarriers. 
While both SBFD and ISAC represent promising directions for future wireless networks \cite{sbfds}, the application of SBFD to ISAC - via allocation of non-overlapping subbands for sensing and communication - remains largely unexplored, and its performance still requires validation through real-world measurements. Under the 3rd Generation Partnership Project (3GPP) regulations, the available bandwidth per frequency band is limited. Compared to TDD-based ISAC, which dedicates a full bandwidth alternately to sensing or communication, SBFD ISAC assigns fewer resources to each function but enables truly continuous operation of both. By employing a single waveform for both tasks, SBFD ISAC achieves a higher degree of signal-domain integration than conventional FDD-based ISAC.

In the proposed SBFD ISAC system, OFDM signals are employed for both sensing and communication, with three non-overlapping subbands allocated - two for sensing and one for communication. System performance is validated through indoor experiments using a testbed comprising three USRP X410 devices, each serving as both transmitter and receiver, centrally managed by a single host server. The setup provides 3 transmit (Tx) and 9 receive (Rx) channels (1 Tx and 3 Rx per USRP). To accommodate the high continuous data flow across multiple channels, each channel operates at 20 MHz bandwidth, ensuring stable data transfer to and from the host where the signals are stored. All channels are configured in the 6.8 GHz band, part of the lower FR3 spectrum (7–24 GHz) that is anticipated to play a key role in future wireless networks. The three USRPs are distributed around the measurement area to form a distributed single-input-multiple-output (SIMO) node system, while a motion capture (MOCAP) system with four cameras provides ground truth references for validating the radar sensing results. 

To validate system feasibility, measurements were conducted under three configurations: (1) SBFD mode with subband allocation at 6.8 GHz, (2) multiband mode with each USRP operating at different center frequencies (6.74 GHz, 6.8 GHz, and 6.86 GHz) with 20 MHz bandwidth, and (3) same-band mode with all USRPs transmitting at 6.8 GHz with 20 MHz bandwidth but without subband separation. In the SBFD configuration, the system achieved a bit-error-rate (BER) of $3.63 \times 10^{-3}$ under non-line-of-sight (NLoS) conditions and a velocity resolution of 0.145 m/s using 1216 symbols for estimation - sufficient for human walking scenarios, where this level of resolution is required. Ground-truth validation was provided by MOCAP data. Root-mean-square error (RMSE) analysis of the velocity estimation confirmed that SBFD ISAC delivers sensing performance comparable to the multiband mode while requiring fewer frequency resources, and significantly outperforms the same-band mode in estimation accuracy. 

\section{SBFD ISAC System Signal Model}
\begin{figure}[htbp]
\centerline{\includegraphics[width=0.25\textwidth]{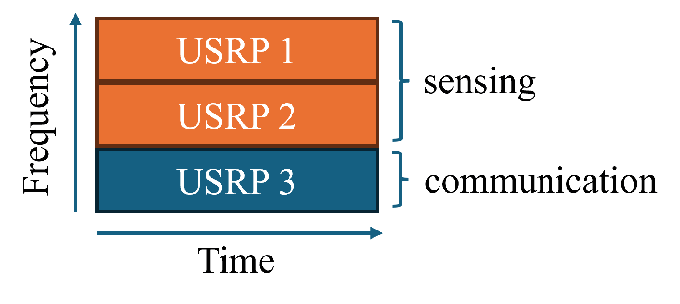}}
\caption{The subband allocation strategy for each USRP}
\label{sbfd}
\end{figure}

The SBFD ISAC system employs OFDM signals with distinct non-overlapping subbands allocated to sensing and communication, enabling simultaneous sensing of the surrounding environment and wireless connectivity. The evaluation testbed comprises three USRP X410 devices, all operating at 6.8 GHz, with each channel configured for a 20 MHz bandwidth. The total bandwidth is divided into three subbands: the first two dedicated to sensing and the third reserved for communication. Each USRP X410 transmits signals containing one of these subbands, as illustrated in Fig.~\ref{sbfd}. 

\subsection{Subcarrier Allocation Strategy Among Distributed Nodes}
The system employs a total of 2048 subcarriers, with allocation defined as follows:
\begin{itemize}
    \item \textbf{USRP 1 (First Sensing Node):}
\begin{align}
\mathcal{S}_1 &= \{65, 66, \ldots, 662\}  \\
\mathcal{G}_1 &= \{1, 2, \ldots, 64\} \cup \{663, 664, \ldots, 2048\}
\end{align}
    \item \textbf{USRP 2 (Second Sensing Node):}
\begin{align}
\mathcal{S}_2 &= \{726, 727, \ldots, 1323\} \\
\mathcal{G}_2 &= \{1, 2, \ldots, 725\} \cup \{1324, 1325, \ldots, 2048\} 
\end{align}
    \item \textbf{USRP 3 (Communication Node):}
\begin{align}
\mathcal{S}_3 &= \{1387, 1388, \ldots, 1984\}  \\
\mathcal{G}_3 &= \{1, 2, \ldots, 1386\} \cup \{1985, 1986, \ldots, 2048\} 
\end{align}
\end{itemize}
Here, $\mathcal{S}_1$, $\mathcal{S}_2$, and $\mathcal{S}_3$ denote the indices of active subcarriers, while $\mathcal{G}_1$, $\mathcal{G}_2$, and $\mathcal{G}_3$ represent the guard subcarriers for each USRP’s OFDM signal. In total, 1794 out of 2048 subcarriers are active across the SBFD ISAC system. This allocation determines the sensing-communication trade-off: each function's performance scales with its allocated bandwidth. Since each receive channel processes the full 20~MHz bandwidth, the inactive guard subcarriers provide spectral isolation between adjacent subbands, mitigating inter-subband leakage and ensuring reliable separation of sensing and communication signals at the receiver.
{The choice of 2048 subcarriers is motivated by two factors: (i) alignment with practical OFDM numerologies standardized in 3GPP, which commonly adopt FFT sizes of 1024, 2048, or 4096 to facilitate efficient implementation; and (ii) compatibility with the USRP~X410 hardware and host-side processing capabilities, balancing subcarrier frequency resolution with real-time I/O throughput constraints. Guard bands are deliberately inserted between subbands to minimize mutual interference, accounting for front-end imperfections such as filter roll-off and synchronization offsets. Furthermore, a pilot spacing of 20 subcarriers is adopted as a practical trade-off between channel estimation accuracy and pilot overhead, enabling robust channel tracking for both sensing and communication functions without incurring excessive spectral cost.}

\subsection{Transmit Signal Model for Distributed Nodes}
The transmit signal models for each of the distributed USRPs are formulated as follows:

\begin{equation} \label{transmit_signal}
s_i(t) = \frac{1}{\sqrt{N}}
\sum_{n \in \mathcal{S}_i} X_i[n] \,
e^{j 2\pi (n-n_0)\Delta f \, t} \;
\mathrm{rect}\!\left(\frac{t+T_g}{T_s+T_g}\right),
\end{equation}
where
\begin{equation*}
X_i[n] =
\begin{cases}
\text{ZC}_i[n], & n \in \mathcal{D}_i, i=1,2 \\[0.5em]
\text{Q}[n], & n \in \mathcal{D}_i, i=3 \\[0.5em]
1, & n \in \mathcal{P}_i, \\[0.5em]
0, & n \notin \mathcal{S}_i.
\end{cases}
\end{equation*}
Here $s_i(t)$ denotes the transmitted signal of the $i$th USRP.
The parameter $n_0=\tfrac{N}{2}+1$ corresponds to the DC-centered bin under 1-based FFT indexing. 
The subcarrier spacing is defined as $\Delta f=1/T_s$, where $T_s$ is the useful OFDM symbol duration (cyclic prefix (CP) excluded), and $T_g$ is the CP duration. The term $X_i[n]$ represents the frequency-domain symbol on the $n$-th subcarrier of the $i$th subband. 
Within each allocated subband $\mathcal{S}_i$, the subsets $\mathcal{D}_i$ and $\mathcal{P}_i$ denote the indices of data and pilot subcarriers, respectively, with $\mathcal{P}_i \cap \mathcal{D}_i = \varnothing$.
In the proposed OFDM signal structure, each active subband contains both data and pilot subcarriers, with pilot tones inserted every 20 subcarriers to balance estimation accuracy and spectral efficiency. For sensing, Zadoff–Chu (ZC) sequences are assigned to the data subcarriers, leveraging their constant-amplitude and zero-correlation zone (ZCZ) properties that reduce peak-to-average power ratio (PAPR) and provide excellent correlation characteristics for range and velocity estimation. These features make ZC sequences particularly suitable for radar-like sensing tasks. For communication, QPSK modulation is employed on data subcarriers, $\text{Q}[n]$ in \eqref{transmit_signal}, as a practical choice, offering a balance between implementation simplicity, spectral efficiency, and robustness.

\subsection{Receive Signal Model of Distributed Nodes}
At the receiver side, each Rx channel of the USRPs captures the full bandwidth, and therefore all subcarriers are received. The received signal at the $k$th USRP can be expressed as
\begin{equation} \label{rx signal}
   r_k(t) = \sum_{i=1}^{3} h_{i,k}(t) \otimes s_i(t) + n_k(t),
\end{equation}
where $r_k(t)$ denotes the {baseband} signal received at USRP~$k$, $h_{i,k}(t)$ is the channel impulse response between transmitter $i$ and receiver $k$, $s_i(t)$ is the transmit signal defined in \eqref{transmit_signal}, and $n_k(t)$ is additive white Gaussian noise. Since the transmit subband signals are pre-defined, the channel responses $h_{i,k}(t)$ can be estimated from the received signals and subsequently exploited for both sensing and communication analysis.

\section{SBFD ISAC Testbed Architecture}
\subsection{Hardware Architecture of the Testbed}
The testbed used for indoor measurements consists of three main hardware components: the host server, the USRP X410, and the antennas. The host server is equipped with a multi-core CPU and 512~GB RAM, providing sufficient capacity for high-speed data buffering. All transmitted and received baseband signals are directly stored in RAM to ensure fast and stable I/O performance.  

The USRP~X410 is a software-defined radio (SDR) platform covering 0–8~GHz and supporting up to four transmit and four receive channels. In our configuration, each device is operated with one transmit channel and three receive channels. To accommodate continuous multi-channel data flow, each channel is configured with a 20~MHz bandwidth, ensuring stable real-time data streaming. The USRPs are connected to the host server through 100~GbE network cards and QSFP28 cables. 

Custom-designed patch antennas are employed for both transmission and reception. Centered at 6.8~GHz, the antennas exhibit a measured 10-dB return-loss bandwidth of approximately 200~MHz (6.7–6.9~GHz) and a realized gain exceeding 12~dBi, confirming that all parameters are within specifications. Each antenna is connected to its corresponding USRP channel via a 1~m coaxial cable. By deploying the USRPs and antennas at distributed locations within the measurement area—enabled by long QSFP28 connections and the independent placement of USRP~X410s—the testbed forms a distributed SIMO node system suitable for SBFD ISAC experimentation. Additional antennas can be ordered to scale the system as required.

\subsection{Software Architecture of the Testbed}

\begin{figure}[htbp]
\centerline{\includegraphics[width=0.35\textwidth]{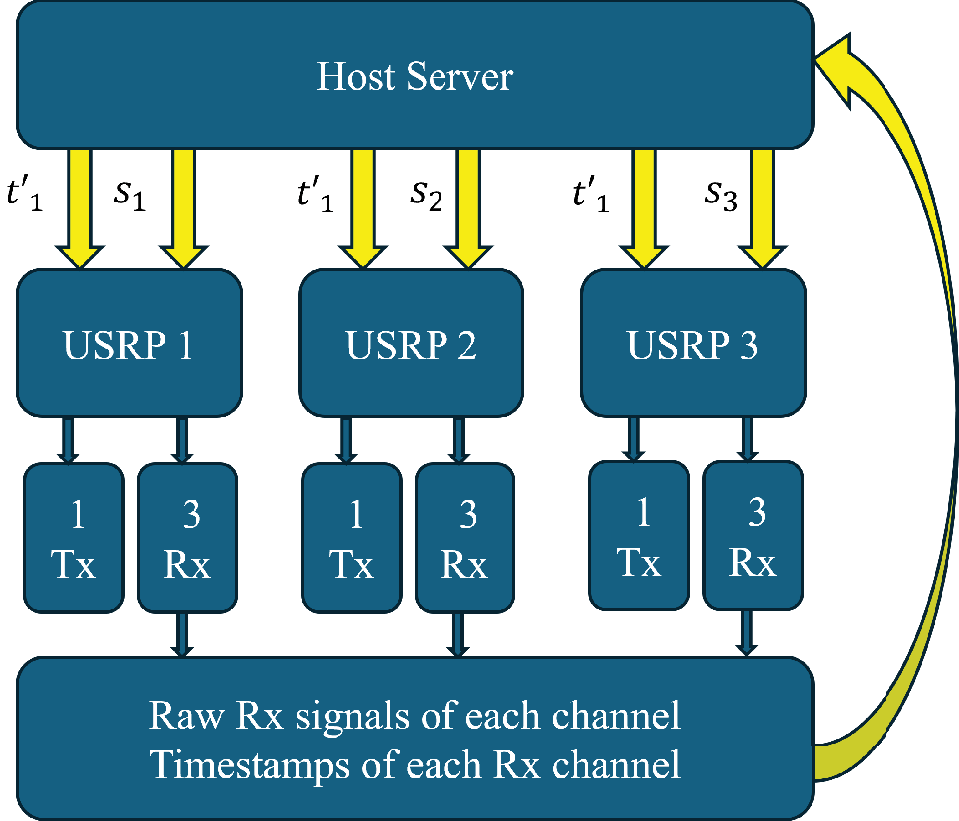}}
\caption{Software architecture of the testbed used for the indoor measurements}
\label{soft}
\end{figure}

A centralized software architecture is adopted to realize the SBFD ISAC system as shown in Fig.~\ref{soft}. The host server performs three main tasks: signal generation, USRP coordination, and data storage.

Signal generation is based on the SBFD ISAC OFDM models described in the previous section and is implemented in MATLAB, with signals stored in complex float-32 format. The generated waveforms $s_1$ and $s_2$ correspond to the sensing signals for USRP~1 and USRP~2, respectively, while $s_3$ is the communication signal for USRP~3. Each waveform occupies its designated subband, while zeros are inserted in the guard bands to preserve spectral isolation between functions.

Synchronization across the three USRP~X410s is achieved by first resetting the clock of each SDR to zero. The host server then provides a common timestamp $t'_1$, which specifies the starting time for the transmit and receive threads on all devices. This procedure reduces the start-time offset between transmit and receive threads across USRPs to below 0.1~ms, which is acceptable given that each channel in the testbed operates independently.

All USRPs capture signals over the full 20~MHz bandwidth, and the raw data is stored in the host server’s RAM. In addition, precise timestamps marking the start of each receive channel are also recorded on the host server, enabling subsequent alignment with ground-truth motion data from the MOCAP system.

\subsection{Transmit OFDM Signal Implementation in the Testbed}
\begin{figure}[htbp]
\centerline{\includegraphics[width=0.45\textwidth]{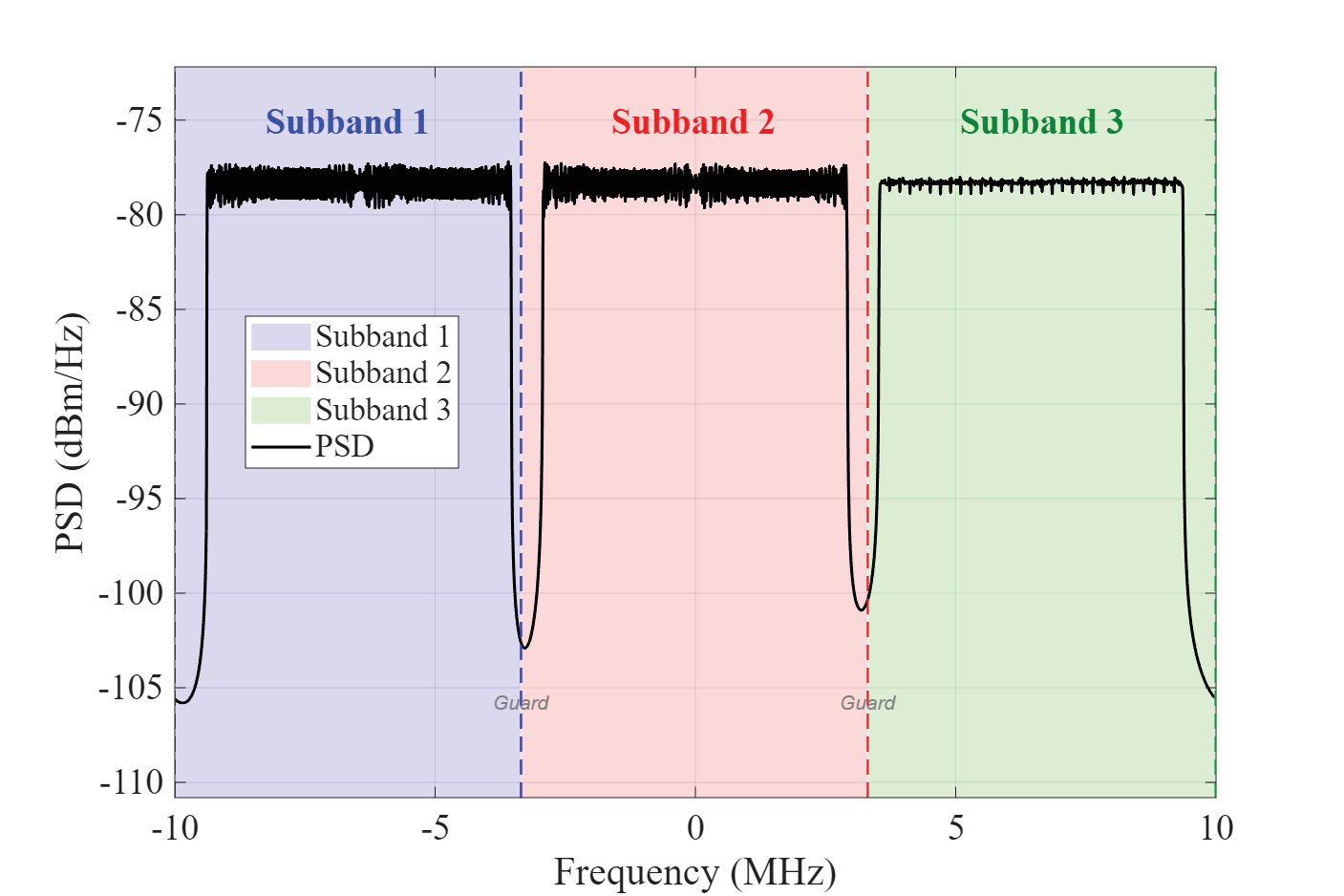}}
\caption{Power spectral density of the combined OFDM signal in SBFD mode, showing three subbands with guard-band separation}
\label{psd}
\end{figure}

In the proposed SBFD ISAC system with distributed SIMO nodes, each USRP transmits a single subband containing a 2048-subcarrier OFDM signal. Based on \eqref{transmit_signal}, the power spectral density (PSD) of the aggregated SBFD OFDM signal is illustrated in Fig.~\ref{psd}. As shown, the three subbands are well separated in the frequency domain, with guard bands inserted to ensure spectral isolation. Subband~1 (blue) and Subband~2 (red) transmit orthogonal Zadoff–Chu sequences for sensing, while Subband~3 (green) carries QPSK-modulated communication data. The OFDM parameters used in the SBFD ISAC system are summarized in Table~\ref{tab:ofdm_params}.

\begin{table}[htbp]
\centering
\caption{OFDM signal parameters used in the SBFD ISAC system}
\label{tab:ofdm_params}
\begin{tabular}{lc}
\hline
\textbf{Parameter} & \textbf{Value} \\
\hline
Center Frequency ($f_c$) & 6.8 GHz \\
Bandwidth ($B$) & 20 MHz \\
Active Subcarriers (per USRP) & 598 \\
Total Symbol Duration ($T_s + T_g$) & 128 $\mu$s \\
\hline
\end{tabular}
\end{table}

\section{Measurement and Validation Methodology}
\subsection{Experimental Configurations with Benchmark Groups}

To evaluate the feasibility of the proposed SBFD ISAC system, we conduct indoor measurements under three different experimental configurations: **SBFD mode**, **multiband mode**, and **same-band mode**. The SBFD mode represents the proposed system, while the other two serve as benchmark groups for comparison.  

All configurations are implemented on the same testbed, consisting of three USRP X410 devices, each using one transmit and three receive channels. The configurations differ in their frequency allocation and signal design strategies:
\begin{itemize}
    \item \textbf{SBFD mode:} The proposed ISAC configuration, where sensing and communication share a single 20~MHz band at 6.8~GHz. Subband allocation and the proposed signal design enable coexistence of both functions.
    \item \textbf{Multiband mode:} Each USRP operates at a distinct center frequency (6.74~GHz, 6.8~GHz, and 6.86~GHz) with 20~MHz bandwidth. All USRPs transmit identical sensing OFDM signals modulated from Zadoff–Chu sequences, fully utilizing their 20~MHz bandwidth for sensing. This minimizes inter-USRP interference and increases sensing bandwidth compared to SBFD mode, but requires a total of 60~MHz spectrum resources.
    \item \textbf{Same-band mode:} A worst-case scenario where all USRPs transmit and receive sensing signals within the same 20~MHz band at 6.8~GHz, without subband separation, leading to severe mutual interference.
\end{itemize}

The key differences between these configurations are summarized in Table~\ref{tab:config}.

\begin{table}[htbp]
\centering
\caption{Comparison of Experimental Configurations}
\label{tab:config}
\begin{tabular}{lccc}
\hline
\textbf{Parameter} & \textbf{SBFD} & \textbf{Multiband} & \textbf{Same-Band} \\
\hline
USRP~1 Center Freq. & 6.8~GHz & 6.74~GHz & 6.8~GHz \\
USRP~2 Center Freq. & 6.8~GHz & 6.8~GHz  & 6.8~GHz \\
USRP~3 Center Freq. & 6.8~GHz & 6.86~GHz & 6.8~GHz \\
Bandwidth per USRP  & 20~MHz  & 20~MHz   & 20~MHz \\
Total Spectrum      & 20~MHz  & 60~MHz   & 20~MHz \\
\hline
USRP~1 Signal & Sensing      & Sensing      & Sensing \\
USRP~2 Signal & Sensing      & Sensing      & Sensing \\
USRP~3 Signal & Communication & Sensing     & Sensing \\
Frequency Separation & Yes & Yes & No \\
\hline
\end{tabular}
\end{table}

\subsection{Measurement Scenario}
\subsubsection{System setup} 
\begin{figure}[bhtp]
\centerline{\includegraphics[width=0.4\textwidth]{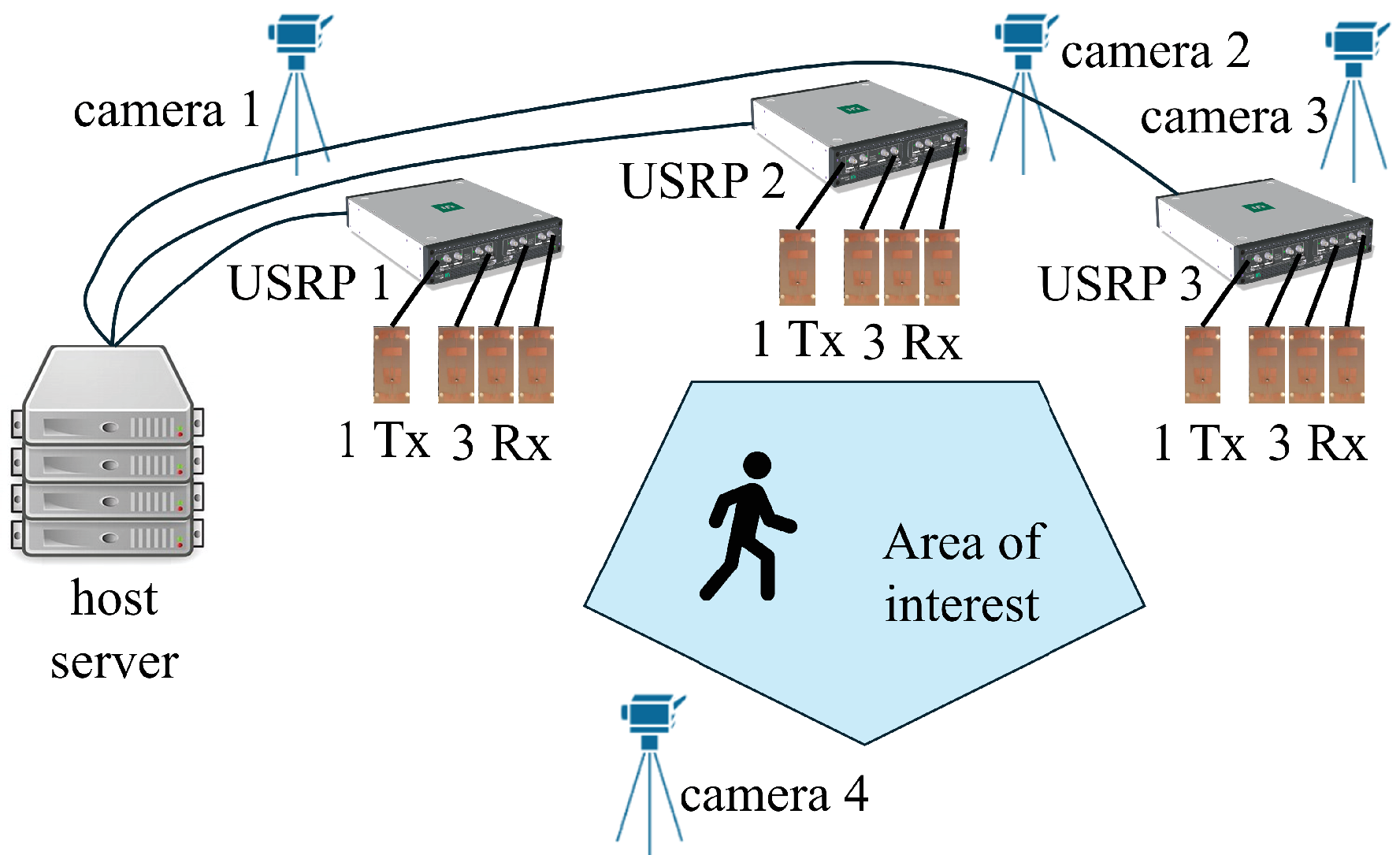}}
\caption{Measurement scenario for the SBFD ISAC testbed}
\label{scenario}
\end{figure}

\begin{figure}[bhtp]
\centerline{\includegraphics[width=0.25\textwidth]{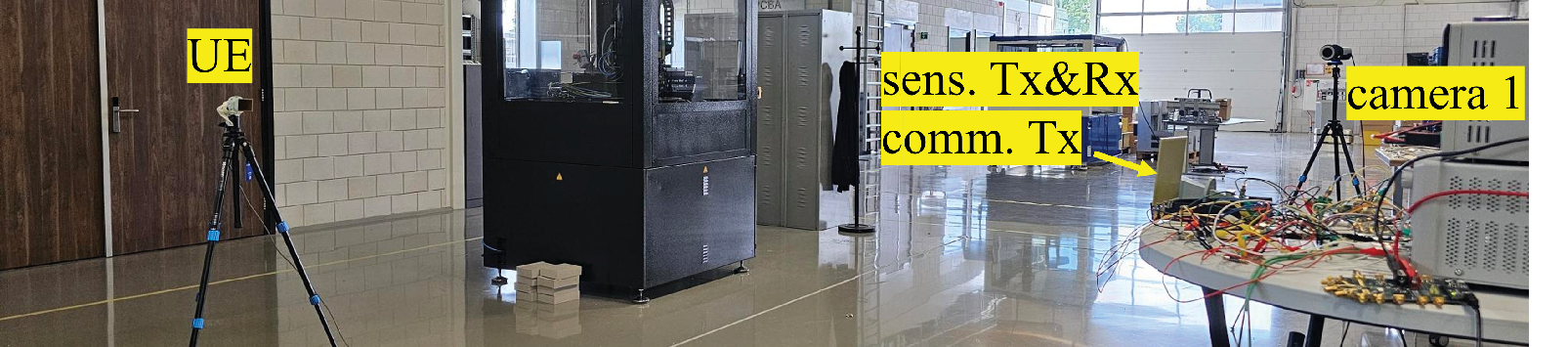}}
\caption{Real measurement set up from the view of the target}
\label{real}
\end{figure}

The measurement scenario used to validate the proposed SBFD ISAC system is depicted in Fig.~\ref{scenario}. As shown, three USRP X410 devices are connected to a common host server and arranged around the area of interest with approximately 1.5~m spacing between adjacent units. Each USRP employs one transmit channel to broadcast its designated signal and three receive channels for data collection.  
Directional antennas with 12~dBi gain, operating in the 6.7–6.9~GHz range, are used for both transmission and reception. To mitigate leakage between the co-located transmit and receive antennas on each USRP, RF absorbers are placed between them. The actual measurement setup, viewed from the target perspective, is presented in Fig.~\ref{real}.

\subsubsection{Coverage Area Validation} Prior to the measurements, the coverage area is validated through real-time monitoring with GNU Radio. A human subject continuously moves within the intended measurement region while the \texttt{QT GUI Frequency Sink} block within GNU Radio monitors received signal power on each USRP channel. Variations in the spectral plots correspond to reflections from the moving subject, confirming that all locations within the area of interest provide sufficient signal quality for both sensing and communication.  

\subsubsection{Measurement procedure} Six independent measurement runs are performed, each lasting 5~s. During each run, a human target walks through the coverage area along a predefined path at random velocities. The testbed operates in one of the three modes (SBFD, multiband, or same-band) during each run. All raw received signals are recorded for subsequent post-processing of both sensing and communication performance.  

\subsubsection{Signal Processing Configuration} 
For sensing evaluation, monostatic processing is applied to the received signals from Subband~1 (USRP~1) and Subband~2 (USRP~2). A window of 1216 OFDM symbols is used for velocity estimation, yielding a velocity resolution of 0.145~m/s. For communication evaluation, OFDM demodulation and QPSK decoding are performed on Subband~3 signals to compute the bit error rate (BER) and generate constellation diagrams.  

\subsubsection{Ground Truth System} 
Ground truth measurements are obtained using a Qualisys motion capture (MOCAP) system with four cameras, managed by a dedicated host laptop. The cameras are positioned around the measurement area, as shown in Fig.~\ref{scenario}. Reflective markers are attached to each RF antenna and to the moving person (sensing target). The MOCAP system records marker positions with sub-millimeter accuracy at a sampling rate of 100~Hz. From this data, relative range and velocity between the subject and each antenna are derived, enabling direct comparison with USRP testbed estimates. Synchronization between the testbed results and the ground truth is achieved by aligning the MOCAP timestamps, including the host laptop start time, with the sensing data.

\subsection{Synchronization of Testbed and MOCAP}
\begin{figure}[bhtp]
\centerline{\includegraphics[width=0.4\textwidth]{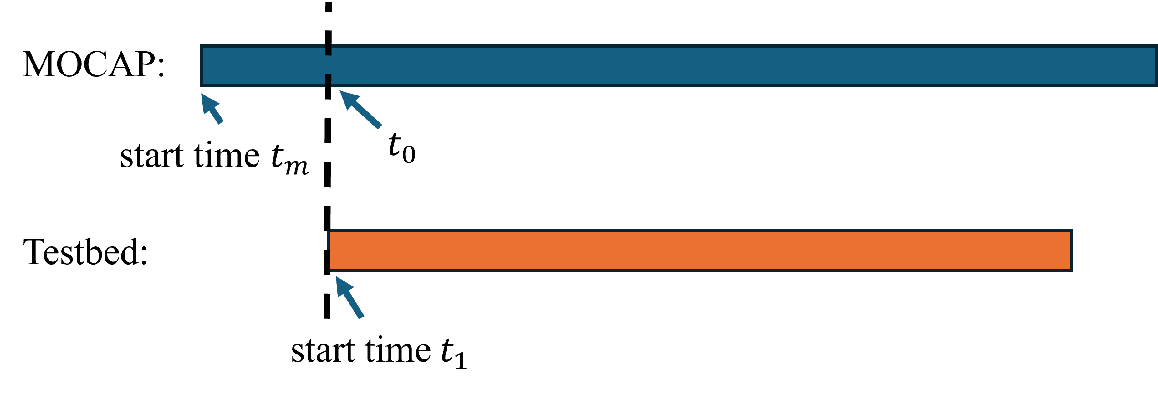}}
\caption{Synchronization method for the CF SBFD ISAC Testbed and the MOCAP system}
\label{sync}
\end{figure}

Synchronization between the testbed and the MOCAP system is achieved through timestamp alignment. The MOCAP system begins recording immediately, whereas the testbed requires a short initialization period. Post-processing compensates for this offset by calculating $t_1 - t_m$, where $t_1$ denotes the testbed start time and $t_m$ the MOCAP start time. As illustrated in Fig.~\ref{sync}, the blue block corresponds to the MOCAP operating period, and the orange block represents the testbed. After alignment, the adjusted MOCAP timeline is defined with a new start time $t_0$.  

\section{Experimental results and Analysis}

\subsection{Received Signal Spectrum}
\begin{figure}[tbhp]
\centerline{\includegraphics[width=0.4\textwidth]{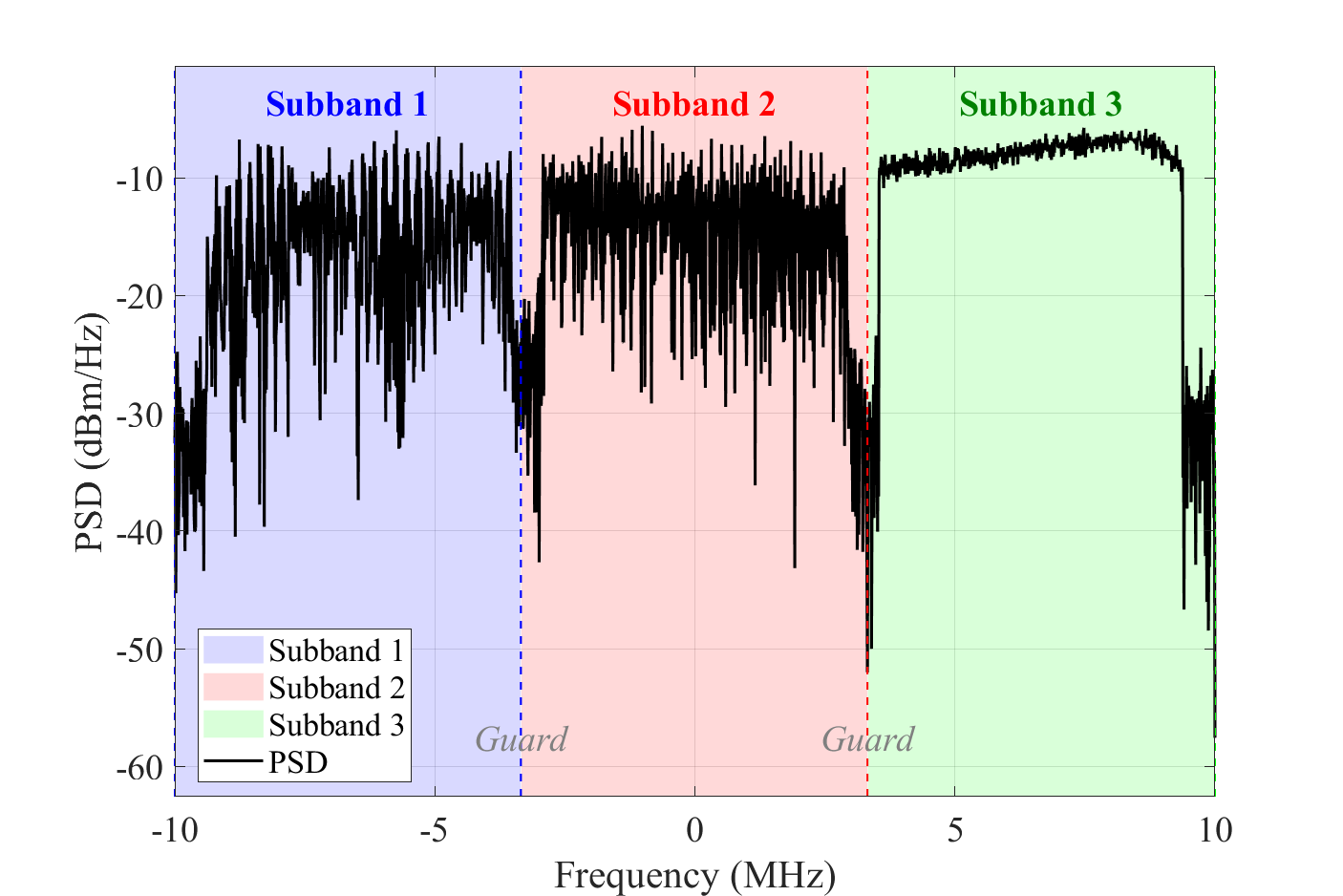}}
\caption{eceived signal spectrum at USRP~2 in SBFD mode, showing all three subbands}
\label{rx}
\end{figure}

Fig.~\ref{rx} presents the spectrum of the received signal from one Rx channel of USRP~2 in SBFD mode. The example confirms that all three subband signals are successfully received. The subbands are clearly separated in the frequency domain, and the receiver channel captures the full 20~MHz bandwidth contributed by the three Tx channels. This separation ensures minimal mutual interference between sensing and communication signals. Subsequent subband-specific signal processing then enables validation of both sensing and communication performance.  

\subsection{Velocity Estimation Performance}
\begin{figure}[bhtp]
\centerline{\includegraphics[width=0.4\textwidth]{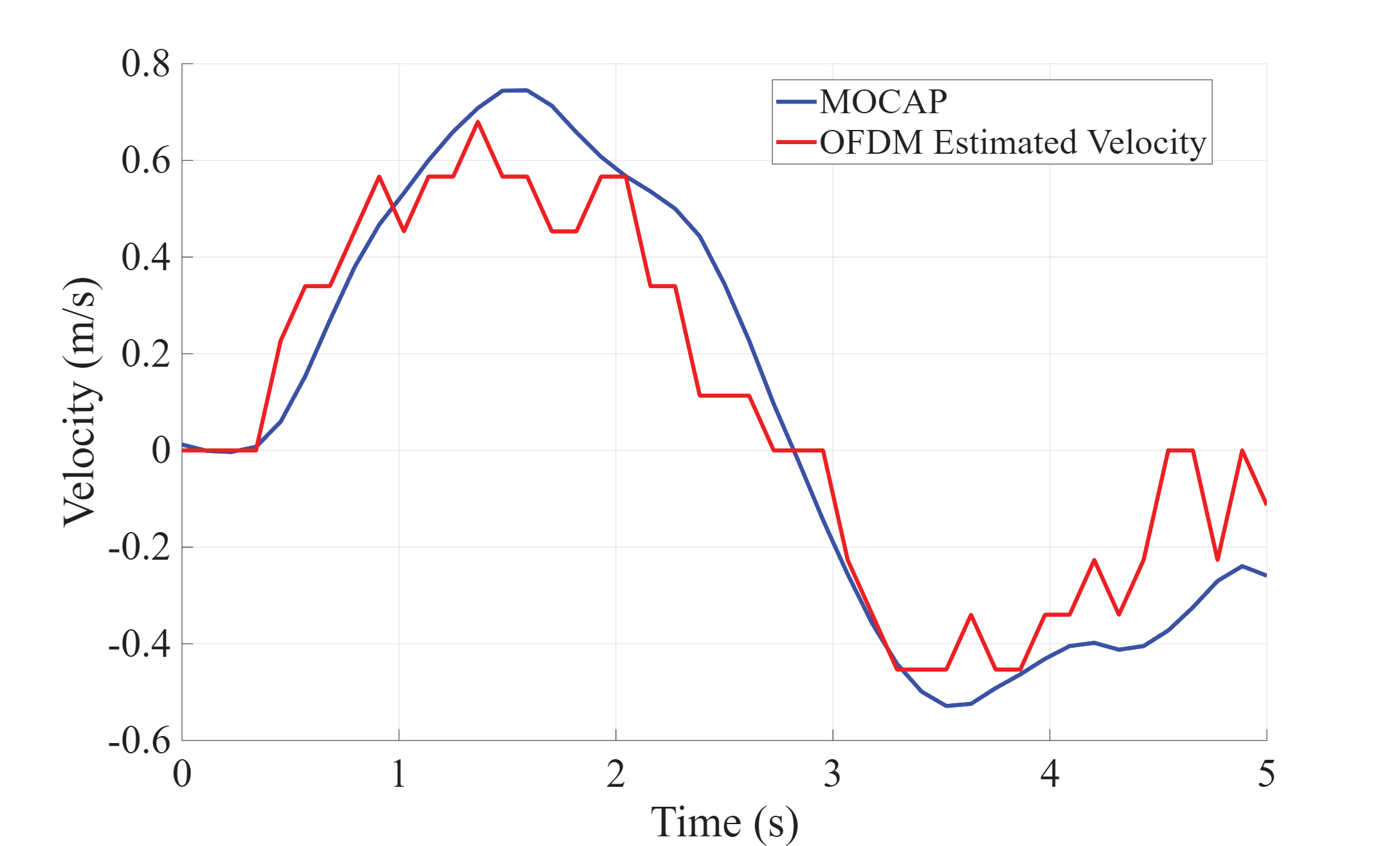}}
\caption{Monostatic velocity estimation from USRP~1 (Subband~1) compared with MOCAP ground truth.}
\label{v1}
\end{figure}

Fig.~\ref{v1} compares the velocity estimation results from a monostatic sensing channel of the SBFD testbed with the MOCAP ground truth data for a representative measurement run. The red curve, obtained from Subband~1 of USRP~1, shows the target’s velocity variation as the person walks through the area of interest. The strong correlation between the estimated velocity trends and the MOCAP measurements validates the sensing capability of the proposed SBFD ISAC system.  

\begin{figure}[bhtp]
\centerline{\includegraphics[width=0.45\textwidth]{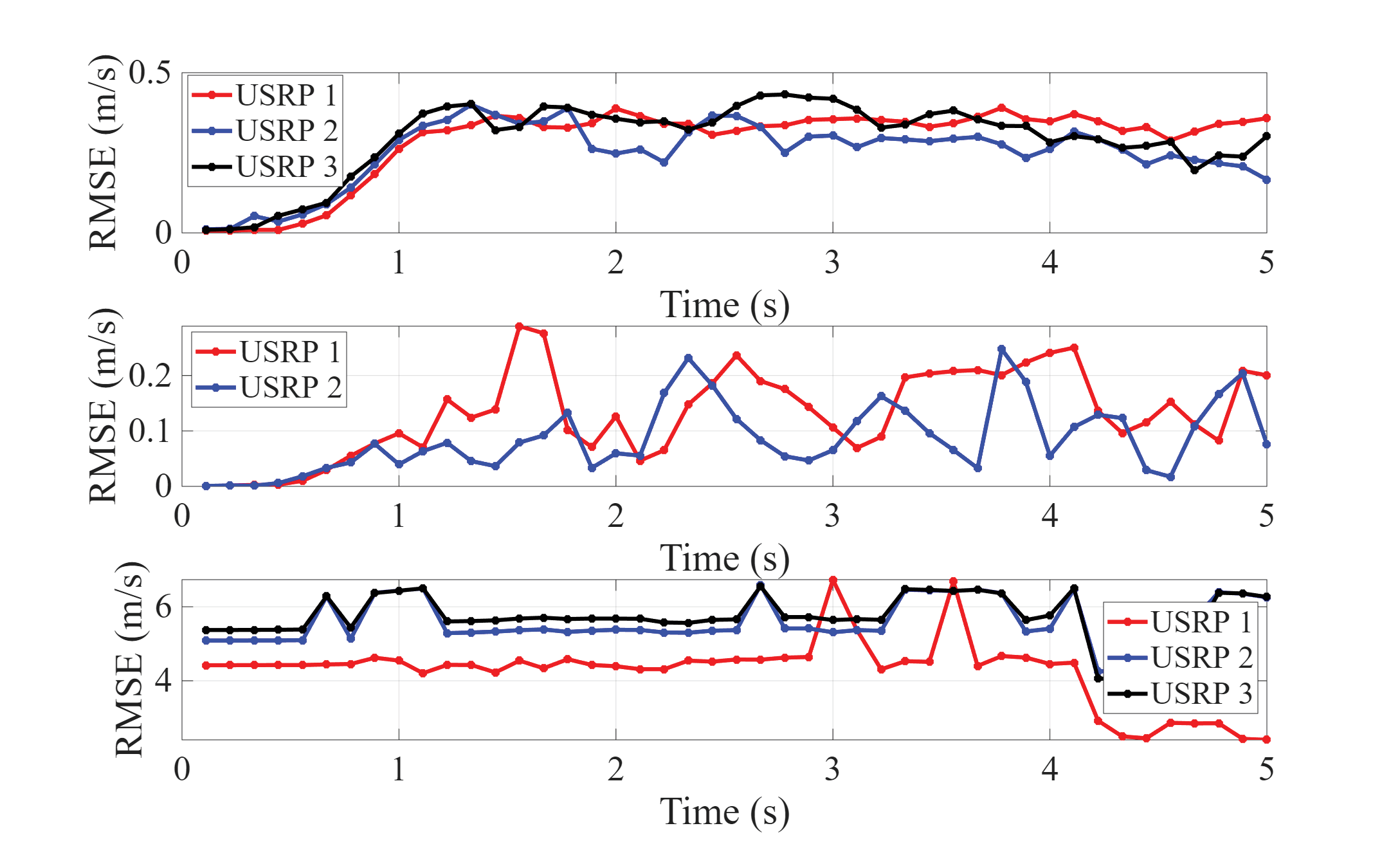}}
\caption{RMSE comparison of velocity estimation for multiband (top), SBFD (middle), and same-band (bottom) configurations}
\label{rmse}
\end{figure}

To further assess the sensing performance of the proposed SBFD system, we quantitatively compare the three modes (SBFD, multiband, and same-band) using the RMSE between the estimated velocity from each USRP and the MOCAP ground truth. The RMSE reflects the average magnitude of estimation errors, with lower values indicating higher accuracy. Fig.~\ref{rmse} presents the comparison results, where the curves represent the average RMSE across six measurement runs at each time step.  
As shown, the multiband mode (top) achieves an RMSE of 0.2–0.4~m/s, while the SBFD mode (middle) demonstrates comparable accuracy with an RMSE of 0.1–0.3~m/s, despite using only 20~MHz bandwidth—one-third of the spectrum resources required by the multiband mode. In contrast, the same-band mode (bottom) suffers severe degradation, with an RMSE of 4–6~m/s, rendering its velocity estimation invalid due to strong mutual interference. Among the three USRPs, USRP~2 exhibits relatively better performance, primarily because its central position in the measurement area provides a wider coverage range.  
These results confirm that the proposed SBFD ISAC system achieves reliable velocity sensing while requiring significantly fewer spectrum resources.

\subsection{Communication Performance}
\begin{figure}[bhtp]
\centerline{\includegraphics[width=0.36\textwidth]{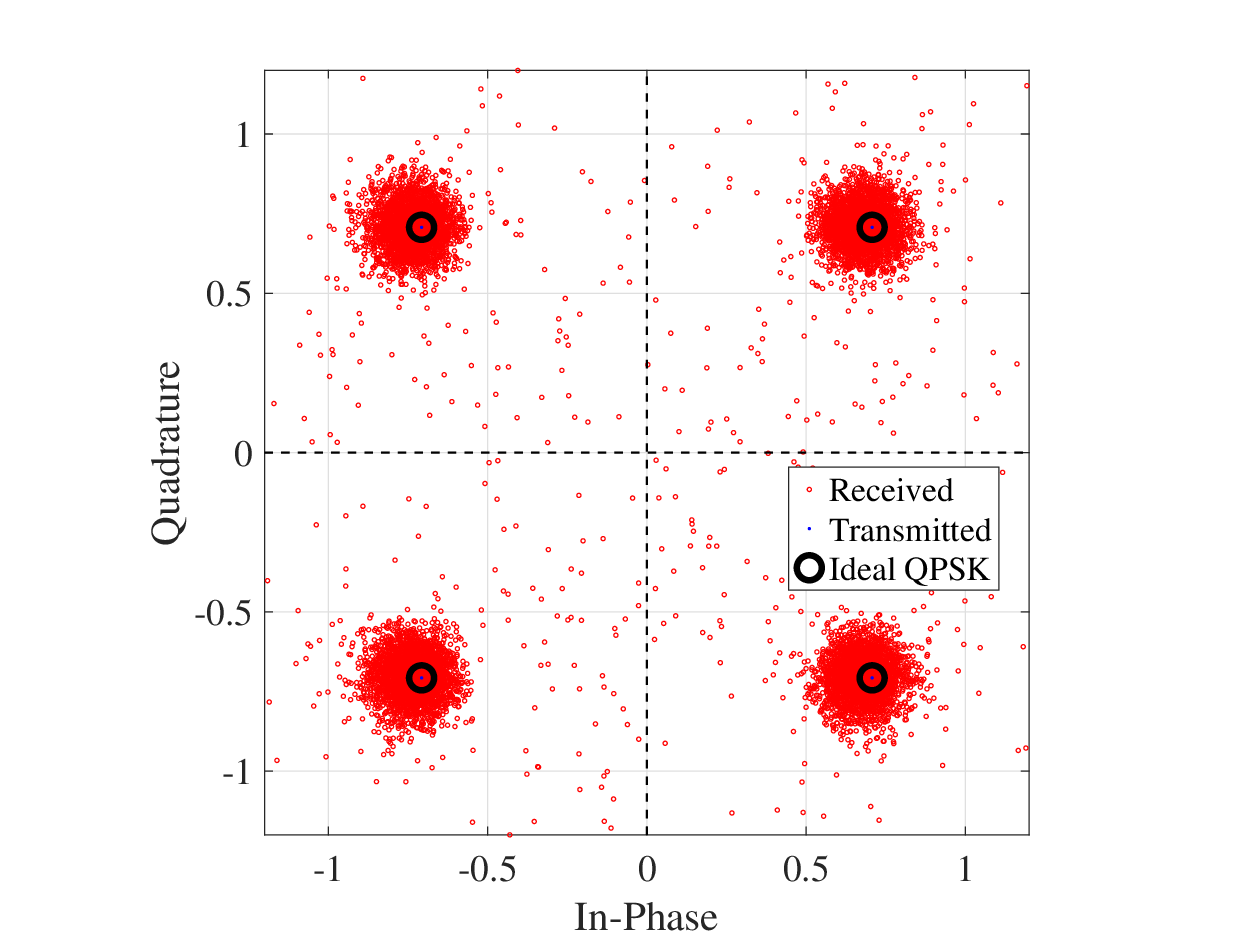}}
\caption{Constellation diagram of a QPSK communication signal received at USRP~3 under NLoS conditions}
\label{cons}
\end{figure}

Since only the SBFD mode includes a communication signal, communication performance is evaluated under this configuration alone. Fig.~\ref{cons} shows the constellation plot of a representative Subband~3 signal received at USRP~3, which operates under NLoS conditions with equalization algorithms. The received symbols (red dots) are clustered around the four ideal QPSK constellation points (black circles), demonstrating effective symbol recovery despite multipath propagation. The measured bit error rate (BER) for this signal is $3.63 \times 10^{-3}$, which falls within the expected range for uncoded QPSK in moderate SNR conditions. This indicates that the proposed system can achieve reliable communication performance even under NLoS propagation.

\section{Conclusions and Future Work}



This paper has presented an SBFD ISAC system with distributed SIMO nodes, validated on an FR3 testbed with USRP~X410s and a MOCAP system. Results show that monostatic sensing achieves a velocity resolution of 0.145~m/s, with SBFD delivering accuracy comparable to multiband (0.1–0.3~m/s vs. 0.2–0.4~m/s) while using only one-third of the spectrum. QPSK communication under NLoS conditions achieved reliable recovery with a BER of $3.63 \times 10^{-3}$. These results confirm that SBFD enables efficient coexistence of sensing and communication with reduced spectral resources, where the sensing and communication performance trade-off is determined by subcarrier allocation rather than interference. Future work will address adaptive subcarrier allocation strategies to deal with the trade-off and multistatic sensing, with the SBFD framework offering a practical step toward FR3 ISAC deployment and 6G standardization.

\section*{Acknowledgment}
This work was supported by the ADA-BEAM project, part of the 6G-BRICKS project funded by the EU Horizon Europe programme. The authors thank KU Leuven ESAT-WAVECORE for providing the testbed infrastructure and measurement facilities.

\end{document}